\documentstyle[12pt]{article}
\begin{document}
\begin{titlepage}
\begin{flushright}
Z\"urich University ZU-TH 1/96\\
\end{flushright}
\vskip 1cm
\begin{center}
{\large\bf MICROLENSING IMPLICATIONS FOR HALO DARK MATTER
\footnote{
Talk presented by Ph. Jetzer at the second workshop on
``The dark side of the Universe: experimental efforts
and theoretical frameworks'' (Rome, 13-14 November 1995).}}\\
\vskip 0.5cm
{\bf Philippe~Jetzer $^2$ and Eduard~Mass\'o $^3$}\\
\vskip 0.5cm
$^2$
Paul Scherrer Institut, Laboratory for Astrophysics,\\
CH-5232 Villigen PSI,\\ and\\
Institute of Theoretical Physics, University of Z\"urich,\\
Winterthurerstrasse 190,
CH-8057 Z\"urich, Switzerland\\
$^3$Departament de F\' \i sica and IFAE, Universitat Aut\`onoma
de Barcelona,\\
 E-08193 Bellaterra, Spain.
\end{center}
\vskip 0.5 cm
\begin{center}
Abstract
\end{center}
\begin{quote}
\baselineskip=12pt

The most accurate way
to get information on the mass of the MACHOs
(Massive Astrophysical Compact Halo Objects)
is to use the method of mass moments.
For the microlensing events detected so far
by the EROS and the MACHO collaborations in the Large Magellanic Cloud
the average mass turns out to be 0.08$M_{\odot}$.
Assuming a spherical standard halo model we find that MACHOs 
contribute about 20\% to the halo dark matter.
The eleven events recorded by OGLE,
mainly during its first two years of operation,
in the galactic bulge lead to an average mass of
0.29$M_{\odot}$, 
whereas forty events detected by MACHO during its first year
give 0.16$M_{\odot}$, thus suggesting
that the lens objects are faint disk stars.
\end{quote}

\end{titlepage}
\newpage
\baselineskip=14pt
\noindent{\bf 1. Introduction}\\

It has been pointed out by Paczy\'nski \cite{kn:Paczynski} that 
microlensing allows the detection of MACHOs in the mass
range \cite{kn:Derujula1}  
$10^{-7} < M/M_{\odot} <  10^{-1}$.
Starting from September 1993 the French collaboration EROS 
\cite{kn:Aubourg} and the American--Australian
collaboration MACHO \cite{kn:Alcock} announced
the detection of
at least six microlensing events discovered by monitoring over several
years millions of stars in the Large Magellanic Cloud (LMC). Moreover,
the Polish-American collaboration OGLE \cite{kn:Udalski} and the 
MACHO team \cite{kn:MACHO} found altogether more than $\sim$ 100  
microlensing events by monitoring stars located in the galactic bulge.
The inferred optical depth for the bulge turns out to be
higher than previously thought.
 
An important issue is 
the determination of the mass of the MACHOs that
acted as gravitational lenses as well as the fraction of halo dark
matter in form of MACHOs.
The most appropriate way to compute the average mass and other
important information is to use
the method of mass moments developed by De R\'ujula et al. 
\cite{kn:Derujula}, which will be briefly presented in section 3.\\ 

\noindent{\bf 2. Most probable mass for a single event}\\

First, we compute the probability $P$ that a microlensing
event of duration T
and maximum amplification $A_{max}$ be produced by a MACHO
of mass $\mu$ (in units of $M_{\odot}$).
Let $d$ be the distance of the MACHO from the line of
sight between the observer and a star in the LMC, t=0 the instant
of closest approach and $v_T$ the MACHO velocity in the transverse
plane. The magnification $A$ as a function of time is calculated using
simple geometry and is given by
\begin{equation}
A(t)=\frac{u^2+2}{u(u^2+4)^{1/2}}~, ~~~{\rm where}~~~
u^2=\frac{d^2+v_T^2 t^2}{R_E^2}~. 
\label{eqno:1}
\end{equation}
$R_E$ is the Einstein radius
which is
$R_E^2=\frac{4GMD}{c^2}x(1-x)=r_E^2 \mu x(1-x)$
with $M = \mu M_{\odot}$ the MACHO mass 
and $D~ (xD)$ the distance
from the observer to the source (to the MACHO). $D = 55$ kpc is the 
distance to the LMC and $r_E=3.17 \times 10^9$~ km.
We use here the definition: $T=R_E/v_T$.
 
We adopt the model of an isothermal spherical halo in which the
normalized MACHO number distribution as a function of $v_T$ is
\begin{equation}
f(v_T) dv_T=\frac{2}{v_H^2} v_T e^{-v_T^2/v^2_H} dv_T~ , \label{eqno:4}
\end{equation}
with $v_H \approx 210$ km/s the velocity dispersion implied by the
rotation curve of our galaxy.
The MACHO number density distribution per unit mass $dn/d\mu$ is
given by
\begin{equation}
\frac{dn}{d\mu}=H(x)\frac{dn_0}{d\mu}=\frac{a^2+R^2_{GC}}
{a^2+R^2_{GC}+D^2x^2-2DR_{GC} x cos \alpha}~\frac{dn_0}{d\mu}, 
\label{eqno:5}
\end{equation}
with $dn_0/d\mu$ the local MACHO mass distribution.
We have assumed that $dn/d\mu$ factorizes in functions of $\mu$ and $x$
\cite{kn:Derujula}.
We take $a = 5.6$ kpc as the galactic ``core'' radius
(our final results do not depend much on the poorly known value of $a$),
$R_{GC} = 8.5$ kpc
our distance from the centre of the galaxy and $\alpha = 82^0$
the angle between the line of sight and the direction of the galactic
centre.
For an experiment monitoring $N_{\star}$ stars during a total 
observation time $t_{obs}$ the number of expected microlensing events is
given by \cite{kn:Derujula,kn:Griest}
\begin{equation}
N_{ev}=\int dN_{ev} =N_{\star} t_{obs}
2Dr_E \int v_T f(v_T) (\mu x(1-x))^{1/2} H(x) \frac{dn_0}{d\mu}
d\mu du_{min} dv_T dx   \label{eqno:6}
\end{equation}
where the integration variable $u_{min}$ is related to $A_{max}$:
$A_{max}=A[u = u_{min}]$.
For a more complete discussion 
in particular on the integration range see \cite{kn:Derujula}.

From eq.(\ref{eqno:6}) with some variable transformation
(see \cite{kn:Jetzer2})
we can define, up to a normalization constant,
the probability $P$ that a microlensing event of
duration $ T$ and  maximum amplification $A_{max}$
be produced by a MACHO of mass $\mu$, that we see first of all
is independent of $A_{max}$ \cite{kn:Jetzer2}
\begin{equation}
P(\mu,T) \propto \frac{\mu^2}{ T^4} \int_0^1 dx (x(1-x))^2 H(x)
exp\left( -\frac{r_E^2 \mu x(1-x)}{v^2_H  T^2} \right) ~. \label{eqno:8}
\end{equation}
We also see that $P(\mu, T)=P(\mu/ T^2)$. The measured values for
$ T$ are listed in the Tables 1 and 2, where 
$\mu_{MP}$ is
the most probable value.
We find that the maximum corresponds to 
$\mu r_E^2/v^2_H T^2=13.0$ \cite{kn:Jetzer2,kn:Jetzer1}. 
The 50\% confidence interval
embraces for the mass $\mu$ approximately
the range $1/3\mu_{MP}$ up to $3 \mu_{MP}$.
Similarly one can compute $P(\mu, T)$ also for the bulge events
(see \cite{kn:Jetzer1}).

\vskip 0.5cm 
{\bf Table 1}: Values of $\mu_{MP}$ (in $M_{\odot}$)
for the six microlensing events detected in the LMC ($A_{i}$
= American-Australian
collaboration events ($i$ = 1,..,4);
$F_1$ and $F_2$
French collaboration events).
For the LMC: $v_H = 210~{\rm km}~{\rm s}^{-1}$ and
$r_E = 3.17 \times 10^9~{\rm km}$.

\begin{center}
\begin{tabular}{|c|c|c|c|c|c|c|}\hline
  & $A_{1}$ & $A_{2}$ & $A_3$ & $A_4$ & $F_1$ & $F_2$  \\
\hline
$ T$ (days) & 16.9 & 9 & 14 & 21.5 & 27 & 30 \\
\hline
$\tau (\equiv \frac{v_H}{r_E} T)$ & 0.097 &
0.052 & 0.08 & 0.123 & 0.154 & 0.172 \\
\hline
$\mu_{MP}$ & 0.12 & 0.03 & 0.08 & 0.20 & 0.31 & 0.38 \\
\hline
\end{tabular}
\end{center} 

\vskip 0.2cm
{\bf Table 2}: Values of
$\mu_{MP}$ (in $M_{\odot}$) as
obtained by the corresponding $P(\mu, T)$ for 
eleven microlensing events detected by OGLE
in the galactic bulge \cite{kn:Jetzer1}. 
($v_H = 30~{\rm km}~{\rm s}^{-1}$ and
$r_E = 1.25 \times 10^9~{\rm km}$.)

\begin{center}
\begin{tabular}{|c|c|c|c|c|c|c|c|c|c|c|c|}\hline
  & 1 & 2 & 3 & 4 & 5 & 6 & 7 & 8 & 9 & 10 & 11\\
\hline
$ T$ & 25.9 & 45 & 10.7 & 14 &12.4& 8.4& 49.5&18.7&61.6&12&20.9 \\
\hline
$\tau$ & 0.054 &
0.093 & 0.022 & 0.029 & 0.026&0.017& 0.103& 0.039& 0.128& 0.025& 0.043\\
\hline
$\mu_{MP}$ & 0.61 & 1.85 & 0.105 & 0.18 &0.14& 0.065& 2.24& 0.32&
3.48 & 0.13 & 0.40 \\
\hline
\end{tabular}
\end{center}
\vskip 0.2cm

\noindent{\bf 3. Mass moment method}\\

A more systematic way to extract information on the masses is to use the
method of mass moments as presented in De R\'ujula et al. 
\cite{kn:Derujula}. The mass moments $<\mu^m>$ are defined as
\begin{equation}
<\mu^m>=\int d\mu~ \epsilon_n(\mu)~ 
\frac{dn_0}{d\mu}\mu^m~. \label{eqno:10}
\end{equation}
$<\mu^m>$ is related to $<\tau^n>=\sum_{events} \tau^n$,
with $\tau \equiv (v_H/r_E) T$, as constructed
from the observations and which can also be computed as follows
\begin{equation}
<\tau^n>=\int dN_{ev}~ \epsilon_n(\mu)~
\tau^n=V u_{TH} \Gamma(2-m) \widehat H(m) <\mu^m>~,
\label{eqno:11}
\end{equation}
with $m \equiv (n+1)/2$ and
\begin{equation}
V \equiv 2 N_{\star} t_{obs}~ D~ r_E~ v_H=2.4 \times 10^3~ pc^3~ 
\frac{N_{\star} ~t_{obs}}{10^6~ {\rm stars/year} }~, \label{eqno:12}
\end{equation}
\begin{equation}
\Gamma(2-m) \equiv \int_0^{\infty} \left(\frac{v_T}{v_H}\right)^{1-n}
f(v_T) dv_T~,
\label{eqno:13}
\end{equation}
\begin{equation}
\widehat H(m) \equiv \int_0^1 (x(1-x))^m H(x) dx~.  \label{eqno:14}
\end{equation}
The efficiency $\epsilon_n(\mu)$ is determined as follows 
(see \cite{kn:Derujula})
\begin{equation}
\epsilon_n(\mu) \equiv \frac{\int d N^{\star}_{ev}(\bar\mu)~ 
\epsilon(T)~ \tau^n}
{\int d N^{\star}_{ev}(\bar\mu)~ \tau^n}~, \label{eqno:15}
\end{equation}
where $d N^{\star}_{ev}(\bar\mu)$ is defined as $d N_{ev}$ 
in eq.(\ref{eqno:6}) with
the MACHO mass distribution concentrated at a fixed mass
$\bar\mu$: $dn_0/d\mu=n_0~ \delta(\mu-\bar\mu)/\mu$. 
In Fig.1 we show the experimental detection efficiency $\epsilon(T)$
of the MACHO experiment when looking to the LMC \cite{kn:Alcock1}. 
In Fig.2 we plot the corresponding $\epsilon_0(\mu)$ as calculated from 
eq.(\ref{eqno:15}). This function indicates how efficient is the
experiment to detect a MACHO with a given mass $M = \mu M_{\odot}$. 

A mass moment $< \mu^m >$ is thus related to 
$< \tau^n >$ as given from the measured values 
of $T$ in a microlensing experiment by
\begin{equation}
< \mu^m > = \frac{< \tau^n >}{V u_{TH} \Gamma(2-m) \hat H(m)}~.
\label{eqno:16}
\end{equation}

The mean local density of MACHOs (number per cubic parsec)
is $<\mu^0>$. The average local mass density in MACHOs is
$<\mu^1>$ solar masses per cubic parsec.  
The  mean MACHO mass, which we get from
the six events detected so far toward the LMC, is \cite{kn:Jetzer1}
\begin{equation}
\frac{<\mu^1>}{<\mu^0>}=0.08~M_{\odot}~.
\label{eqno:aa}
\end{equation}
(To obtain this result we used the values of $\tau$
as reported in Table 1, whereas $\Gamma(1)\widehat H(1)=0.0362$ and
$\Gamma(2)\widehat H(0)=0.280$ as
plotted in figure 6 of ref. \cite{kn:Derujula}).

The mean MACHO mass, which one gets from
the eleven events of OGLE in the galactic bulge
is $\sim 0.29 M_{\odot}$ \cite{kn:Jetzer1}.
From the 40 events discovered \footnote{We considered
only the events used by the MACHO team to infer the optical
depth without the double lens event.} 
during the first year of operation
by the MACHO team \cite{kn:MACHO} we get an average value
of 0.16$M_{\odot}$.
The lower value inferred from the MACHO data is due to the fact
that the efficiency for the short duration events ($\sim$ some days)
is substantially higher for the MACHO experiment than for the
OGLE one. 
The above average values for the mass
suggests that the lens are faint disk
stars. 

The resulting mass depends obviously to some extent on the parameters
used to describe the halo (or the galactic centre
respectively). In order to check this
dependence we varied the parameters within
their allowed range and found
that the average mass changes at most by $\pm$ 30\%, which shows
that the result is rather robust. 

Another important quantity is the fraction $f$ of the local
dark mass density (the latter one given by $\rho_0$) detected
in the form of MACHOs, which is given by
$f \equiv {M_{\odot}}/{\rho_0} \sim 126~{\rm pc}^3$ $<\mu^1>$.
Using the values given by the MACHO collaboration
for their first year data \cite{kn:Alcock1} (in particular
$u_{TH}=0.83$ corresponding to $A > 1.5$ and
an effective exposure $N_{\star} t_{obs}$
of $\sim 2 \times 10^6$ star-years for 
the observed range of the event duration $T$ between 10 - 20 days)
we find $f \sim 0.2$, which compares quite well
with the corresponding value ($f=0.19^{+0.16}_{-0.10}$) obtained 
by the MACHO group in a different way.

Once several moments $< \mu^m >$ are known one can
get information on the mass distribution $dn_0/d\mu$. However,
since at present only few events toward the LMC are at disposal the 
different moments (especially the higher ones) can only 
be determined approximately.
Instead, we can make the ansatz $dn_0/d\mu=a \mu^{-\alpha}$.
Knowing, for instance, $<\mu^1>$ and $<\mu^0>$ (as well
as $\epsilon_1(\mu)$ and $\epsilon_{-1}(\mu)$ from eq.(\ref{eqno:15}))
we can determine $a$ and $\alpha$.
The solution for $a$ and $\alpha$ is acceptable only if we get
the same values using other moments, such as e.g. $<\mu^{1.5}>$.
Remarkably, we find that $a \simeq 6.5 \times 10^{-4}$ and $\alpha 
\simeq 2$ is a consistent solution.
Moreover, from the relation
\begin{equation}
\int^{\sim 0.1}_{M_{min}} \frac{dn_0}{d\mu} \mu d\mu = f \rho_0 
\end{equation}
with the above values for $a$, $\alpha$ and $f \simeq 0.2$
it follows that $M_{min} \sim 10^{-2} M_{\odot}$.
Obvioulsy these results have to be considered as preliminary
and as an illustration of how one can get useful information
with the mass moment method.
Once more data are available it will also be possible
to determine other important quantities
such as the statistical error in eq. (\ref{eqno:aa}) .

Nevertheless, the results obtained so far
are already of interest and it is clear that in a few years
it will be possible to draw more firm conclusions.

\vskip 3truecm

\noindent
{\Large \bf Figure Captions}\\

{\bf 1.} $\epsilon(T)$ as given by the MACHO collaboration.\\

{\bf 2.} $\epsilon_0(\mu)$ as one gets with eq.(\ref{eqno:15}).

\end{document}